\documentclass[letterpaper,english,english,prd,preprintnumbers,twocolumn,nofootinbib,showpacs,shortbibliography,notitlepage]{revtex4-1}
\usepackage[T1]{fontenc}
\usepackage[latin9]{inputenc}
\usepackage{amsmath}
\usepackage{amssymb}

\makeatletter

\usepackage{verbatim}
\usepackage{graphicx}
\usepackage{babel}
\makeatother

\usepackage{babel}
\begin{document}
\global\long\def\edge#1{\left.#1\right|}
\global\long\def\d{\mathrm{d}}
\global\long\def\order#1{\mathcal{O}\left(#1\right)}
\def\Vud{V_\text{ud}}
\def\Vus{V_\text{us}}
\def\Vub{V_\text{ub}}
\def\BR{\text{BR}}
\def\sig{$\sigma$}
\def\second{\text{s}}
\selectlanguage{english}

\preprint{Alberta Thy 1-18}

\title{Neutron Lifetime and Axial Coupling Connection }

\author{Andrzej Czarnecki }
\affiliation{Department of Physics, University of Alberta, Edmonton, Alberta, Canada T6G 2E1}
\author{William J.~Marciano}
\affiliation{Department of Physics, Brookhaven National Laboratory,
  Upton, New York 11973, USA}

\author{Alberto Sirlin}
\affiliation{Department of Physics, New York University,\\
    726 Broadway, New York, New York 10003, USA}

\begin{abstract}
  Experimental studies of neutron decay, $n\to pe\bar\nu$, exhibit two anomalies.  The first is a 8.6(2.1)s, roughly $4\sigma$ difference between the average beam measured neutron lifetime, $\tau_n^\text{beam}=888.0(2.0)$s, and the more precise average trapped ultra cold neutron determination, $\tau_n^\text{trap}=879.4(6)$s. The second is a $5\sigma$ difference between the pre2002 average axial coupling, $g_A$, as measured in neutron decay asymmetries $g_A^\text{pre2002}=1.2637(21)$, and the more recent, post2002, average $g_A^\text{post2002}=1.2755(11)$, where, following the UCNA collaboration division, experiments are classified by the date of their most recent result. In this study, we correlate those $\tau_n$ and $g_A$ values using a (slightly) updated relation $\tau_n(1+3g_A^2)=5172.0(1.1)$s. Consistency with that relation and better precision suggest $\tau_n^\text{favored}=879.4(6)$s and $g_A^\text{favored}=1.2755(11)$ as preferred values for those parameters.  Comparisons of $g_A^\text{favored}$ with recent lattice QCD and muonic hydrogen capture results are made.  A general constraint on exotic neutron decay branching ratios, $<0.27\%$, is discussed and applied to a recently proposed solution to the neutron lifetime puzzle.
\end{abstract}
\maketitle

The neutron lifetime, $\tau_n$, and its axial-current coupling,
$g_A=G_A/G_V$, are important weak interaction parameters used in
nuclear, particle and astro physics as well as cosmology
\cite{Abele:2008zz,Dubbers:2011ns,Wietfeldt:2011suo,Severijns:2011zz,Ivanov:2012qe,greene:2016aa}. Employed
together, they can determine the quark mixing matrix element $\Vud$,
at a level that could eventually become competitive with the current
super-allowed Fermi transition nuclear beta decay method for
determining $\Vud$ \cite{Patrignani:2016xqp} and constraining ``New
Physics'' via CKM unitarity $|\Vud|^2+|\Vus|^2+|\Vub|^2 =1$. Neutron
decays have the advantage of no nuclear physics uncertainties
\cite{Abele:2003ya}.

On its own, $g_A$ provides necessary input for the Goldberger-Treiman 
relation, Bjorken Sum Rule, solar and reactor neutrino fluxes, 
neutrino-nucleon quasi-elastic scattering cross-sections, muon capture rates 
and various other weak interaction phenomena. An area of particular 
importance is the dependence of primordial nucleosynthesis and Cosmic 
Microwave Background Anisotropies on $\tau_n$ and $g_A$
\cite{Mathews:2004kc,Pitrou:2018cgg}.  

Despite their central role in weak interaction phenomenology, $\tau_n$
and $g_A$ values have changed, sometimes dramatically, with
time. Indeed, the accepted $\tau_n$ has decreased over the Particle
Data Group (PDG) lifespan from about
$1000\second \to 932\second \to 917\second \to 896\second \to
886\second$ while over a similar time span $g_A$ has increased from
roughly $1.20\to 1.23\to 1.25\to 1.26\to 1.27$.  The correlated
movement with time of $\tau_n$ and $g_A$ is nicely illustrated in the
introduction figures of ref.~\cite{Patrignani:2016xqp}.  As we shall
argue in this paper, further change in both quantities appears to be
in progress. Although the most precise $\tau_n$ and $g_A$ experimental
measurements have generally been carried out independently of one
another, prevailing values at a given time were known to be correlated
through the relationship $\tau_n(1+3g_A^2)=\text{constant}$, with the
constant determined by the Standard Model (SM) neutron decay rate
prediction. Thus, $\tau_n$ and $g_A$ experimental values can be
expected to move together. Here, we review and update
(very slightly) the origin, uncertainty and status of that constant,
by updating the inputs, checking the analysis, and assigning an
uncertainty to the theory prediction. 

Currently, there are two competing values for $\tau_n$ and two for
$g_A$ (see Table \ref{tab:aver}).  Although the values in each set are generally
averaged by the PDG with errors increased by a scale factor based on
the $\chi^2$, we keep them separate. The
average beam measurements $\tau_n^\text{beam}=888.0(2.0)$s differ  by about
4\sig\ 
from the newer
more precise ultra cold trapped neutron average
$\tau_n^\text{trap}=879.4(6)$s. That difference is sometimes referred to as the neutron
lifetime puzzle, enigma or problem. Similarly,  an earlier set
of $g_A$ measurements labeled pre2002 averages to
$g_A^\text{pre2002}=1.2637(21)$ while determinations completed after
2002, labeled post2002, average to $g_A^\text{post2002}=1.2755(11)$, a
5\sig\ difference, even more pronounced than the neutron
lifetime problem.  A notable difference \cite{Abele:2002wc,Abele:2008zz}, between
pre and post 2002 experiments is that the earlier efforts required
larger corrections to the measured asymmetries.  As a result,  those
corrections and their systematic uncertainties may have been more
difficult to properly estimate. The two $g_A$ values are generally PDG
averaged and the uncertainty is increased by a scale factor of
approximately 2,  primarily due to pre2002 $\chi^2$ contributions.
Here, we keep the method dependent $\tau_n$ as well
as the pre and post 2002 $g_A$ values separate and argue in favor of
the more recent values in both cases, because of their better precision
and, more important, their remarkable consistency with our evaluation of the constant in the $\tau_n$-$g_A$ relation
previewed above. On that basis, we will argue that, within the SM,
$\tau_n^\text{favored}=879.4(6)\second$ and
$g_A^\text{favored}=1.2755(11)$ currently represent our recommended
``favored values''. They may be the final word, within
errors.

\begin{table}[htb]
 \caption{Input data used for the $\tau_n^\text{trap}$, $\tau_n^\text{beam}$, 
   $g_A^\text{post2002}$ and $g_A^\text{pre2002}$ averages. Values and 
   methodology were based on PDG2016 but with updates 
   from 
   \protect\cite{Serebrov:2017bzo,Pattie:2017vsj, Ezhov:2014tna,Brown:2017mhw}. The 
   error in $\tau_n^\text{trap}$  average was scaled by a factor of 
   1.5 in accordance with PDG protocol. Statistical and systematic 
   uncertainties were added in quadrature and kept to two significant figures before 
   averaging. Averages have not been sanctioned by the PDG.} 
  \centering
  \begin{tabular}{l @{\hspace{5mm}} l}
\hline 
\hline 
           $\tau_n^\text{trap}$          &     Source     \\
          881.5(0.92)$\second$       & \cite{Serebrov:2017bzo} \\
          877.7(0.76)$\second$       & \cite{Pattie:2017vsj} \\
          878.3(1.9)$\second$ & \cite{Ezhov:2014tna} \\
          880.2(1.2)$\second$       & \cite{Arzumanov:2015tea}\\
          882.5(2.1)$\second$       & \cite{Steyerl:2012zz} \\
          880.7(1.8)$\second$       &\cite{Pichlmaier:2010zz}\\
          878.5(0.76)$\second$       & \cite{Serebrov:2004zf}   \\
          882.6(2.7)$\second$       &\cite{Mampe:1993an}\\
    879.4(6)$\second$    & Average (includes scale factor $S=1.5$)\\
\hline 
\hline 
           $\tau_n^\text{beam}$          &     Source     \\
          887.7(2.2)$\second$     &  \cite{Yue:2013qrc} \\
          889.2(4.9)$\second$     &\cite{Byrne:1996zz}\\
  888.0(2.0)$\second$ & Average  \\
\hline  
\hline  
     $ g_A^\text{post2002}$        &     Source     \\
           1.2772(20)     &    \cite{Brown:2017mhw} \\
           $1.2748^{+13}_{-14}$ &    \cite{Mund:2012fq}\\
           1.2750(160)       & \cite{Schumann:2007qe}\\
       1.2755(11) & Average  \\
\hline
\hline  
     $ g_A^\text{pre2002}$        &     Source     \\
           1.2686(47)       &  \cite{Mostovoi:2001ye} \\
           1.2660(40)       &  \cite{Liaud:1997vu}\\
           1.2594(38)       &  \cite{Erozolimsky:1997wi}\\
           1.2620(50)       &  \cite{Bopp:1986rt}\\
           1.2637(21)  & Average  \\
\hline 
\hline 
 \end{tabular}
  \label{tab:aver}
\end{table}

Relating $\tau_n$ and $g_A$ begins with a very precise 
SM prediction for the total (radiative inclusive) neutron decay
rate.  That inverse lifetime formula includes Fermi Function final
state electron-proton Coulomb interactions, electroweak radiative
corrections (normalized relative to the muon lifetime  \cite{Marciano:2003ci}) and a number of
smaller effects including proton recoil, finite nuclear size
etc. Overall, those corrections are rather large, $> +7$\%.  A very detailed analysis of those corrections was given in the
classic study by Wilkinson \cite{Wilkinson:1982hu}. Later, that
relationship was checked, updated and refined in
\cite{Czarnecki:2004cw} where higher order 
$\mathcal{O}(\alpha^2)$ contributions were properly included. The
radiative corrections uncertainty was reduced in
\cite{Marciano:2005ec}. 

In the SM, the inverse lifetime equation relating $\tau_n$ and $g_A$ is given by \cite{Czarnecki:2004cw}
\begin{equation}
  \label{eq:2}
{1\over \tau_n} = {G_\mu^2|\Vud|^2\over 2\pi^3} m_e^5 \left(1+3g_A^2\right)
\left( 1+\mbox{RC}\right) f,
\end{equation}
where $G_\mu$ is the Fermi constant determined from the muon lifetime
\cite{Kinoshita:1958ru,Berman:1958ti,berman62,Abers:1968zz,vanRitbergen:1998yd,%
  Steinhauser:1999bx,Ferroglia:1999tg,
  Marciano:1999ih,Blokland:2004ye,Pak:2008qt},
$G_\mu=1.1663787(6)\times 10^{-5}\,\text{GeV}^{-2}$, $\Vud$ is the CKM
mixing element generally obtained from super-allowed nuclear beta
decays \cite{Patrignani:2016xqp,Hardy:2016vhg}, RC represents
electroweak radiative corrections
\cite{Sirlin:1967zza,SirlinAustria,Sirlin:1974ni,Sirlin:1978sv,Blin-Stoyle:1970sz,Sirlin:2003ds,Sirlin:2012mh}
which were most recently evaluated \cite{Marciano:2005ec} to be
$+0.03886(38)$ and $f$ is a phase space factor \cite{Wilkinson:1982hu}.  The electroweak radiative
corrections in eq.~\eqref{eq:2} have been factorized to be the same
for vector and axial-vector contributions \cite{Czarnecki:2004cw}.  That prescription
defines $g_A$ as determined by the neutron lifetime. Expressing the
polarized neutron spin-electron correlation coefficient, $A_0(g_A)
=2g_A(1-g_A)/\left(1+3g_A^2\right)$, in terms of that $g_A$ will,
therefore, 
induce small $\order{0.1\%}$ radiative corrections \cite{Shann:1971fz} along with the
$\order{1\%}$  residual Coulomb, recoil and weak magnetism corrections
to the measured asymmetry that must be corrected for before extracting
$g_A$ \cite{Ivanov:2012qe, Wilkinson:1982hu}.

Employing masses \cite{Patrignani:2016xqp} (with highly correlated
uncertainties due to atomic mass units to MeV translation)
$m_n = 939.5654133(58)\text{ MeV}$,
$m_p = 938.2720813(58)\text{ MeV}$, and
$m_e = 0.5109989461(31)\text{ MeV}$ leads to $f= 1.6887(1)$
\cite{Wilkinson:1982hu,Czarnecki:2004cw}, where we have redone the
numerical evaluation of Wilkinson's perturbative analysis and employed
a conservative error consistent with his assessment
\cite{Wilkinson:1998hx}.  Using the above input parameters, but
keeping $\Vud$, $\tau_n$ and $g_A$ arbitrary, produces the SM master
formula
\begin{equation}
  \label{eq:1}
  |\Vud|^2 \tau_n(1+3g_A^2)= 4908.6(1.9)\second
\end{equation}
where the uncertainty comes primarily from the RC. That formula can be
used to determine $\Vud$ from independent experimental measurements of
$\tau_n$ and $g_A$. Future experiments optimistically hope to
eventually reach $\pm 0.01$\% sensitivity for those input
parameters. At that level, the RC theory uncertainty will be dominant.

Our intention is to correlate $\tau_n$ and $g_A$ rather than determine
$\Vud$. To that end, we employ the super-allowed $0^+\to 0^+$ nuclear
transitions current best value $\Vud =0.97420(10)(18)_\text{RC}$, a
value consistent with CKM unitarity \cite{Patrignani:2016xqp}, where
the first error $(10)$ results from experiment, nuclear structure and
nucleus dependent radiative corrections, while the second error
$(18) _\text{RC}$ represents universal radiative corrections common to
both neutron and nuclear beta decays.  Importantly, the RC error in
$|\Vud|^2$ and in eq.~\eqref{eq:2} are anticorrelated and effectively
cancel. For that reason, one finds the following very precise relation
between $\tau_n$ and $g_A$
\begin{equation}
  \label{eq:5}
\tau_n(1+3g_A^2) = 5172.0(1.1)\second
\end{equation}
where the uncertainty stems primarily from nuclear and experimental
uncertainties in $\Vud$. That connection allows one to translate
between $\tau_n$ and $g_A$ with high precision and thereby test their
mutual consistency. 

In that way, lifetime and axial-charge measurements can
be directly compared or for some purposes even averaged. Toward that
end, it is useful to divide the lifetime averages into
trap, which includes bottle and magnetic confinement trap experiments,
and beam measurements, the two areas of disagreement. Similarly,
following the classification introduced by the UCNA Collaboration
\cite{Brown:2017mhw}, asymmetry values of $g_A$ naturally separate
into pre2002 and post2002, where 2002 represents the year when larger
values of $g_A$, seen earlier, were confirmed with improved errors
\cite{Abele:2002wc,Abele:1997yn}. Experiments are arranged by the year
of their last result (see Table \ref{tab:aver}).  The post2002
measurements of $g_A$ tended to have larger central values and better
controlled systematics.  That approach leads to the following direct
and indirect averages, connected by arrows representing the
relationship in eq.~\eqref{eq:5},
\begin{align}
  \label{eq:7}
\tau_n^\text{trap}=879.4(6) \second &\to  g_A=1.2756(5) 
\\
\tau_n^\text{beam} =888.0(2.0) \second &\to  g_A=1.2681(17)\\
g_A^\text{post2002} =1.2755(11) &\to  \tau_n=879.5(1.3)\second\\ 
g_A^\text{pre2002} =1.2637(21) &\to  \tau_n=893.1(2.4)\second
\end{align}
One notices that $\tau_n^\text{trap}$ and $g_A^\text{post2002}$
provide the most precise direct and indirect lifetimes, respectively,
and they are remarkably consistent.
Those features are illustrated in Fig.~6 of Ref.~\cite{Brown:2017mhw}. 
That agreement
is exactly the type of consistency one expects of the true parameters.
On the other hand, the beam and pre2002 $g_A$ determined lifetimes
disagree with those more precise values and are not particularly
consistent with one another. 
Because of their better precision and
relationship consistency, we refer to the trap lifetime and post2002 $g_A$ as
our favored values,
\begin{align}
  \label{eq:6a}
  \tau_n^\text{favored} & =879.4(6)\second\\
 \label{eq:6b}
  g_A^\text{favored} & = 1.2755(11).
\end{align}
These favored experimental averages, in conjunction with the indirect
determination of $g_A$ in Eq.~\eqref{eq:7}, provide standards for comparison with
future lifetime and asymmetry measurements which will aim at the long
term goal of $0.01\%$ precision in $\tau_n$ and $g_A$. 
Our current favored values in
Eqs.~\eqref{eq:6a} and \eqref{eq:6b} should be compared with our
updates of the 2016 PDG averages based on recent results
\cite{Serebrov:2017bzo,Pattie:2017vsj, Ezhov:2014tna, Brown:2017mhw} in
Table \ref{tab:aver},
\begin{align}
  \label{eq:6}
\tau_n^\text{update16}&=879.7(8) \second \quad\text{(with scale factor $S=2$)}
\\
 \label{eq:15n}
 g_A^\text{update16} &= 1.2731(23) \quad\text{(with  $S=2.3$)}.
\end{align}
Those updates are consistent with our preferred values in Eqs.~(\ref{eq:6a},\ref{eq:6b});
but, have larger errors due to scale factors that represent
inconsistencies in the experiments averaged. They are useful as a
conservative perspective on the current $\tau_n$ and $g_A$ situation. 

Regarding our neglect of $\tau_n^\text{beam}$ and $g_A^\text{pre2002}$
in deriving our favored values, we make the following
observations. $\tau_n^\text{beam}$ differs from $\tau_n^\text{trap}$
by about 4\sig\ and $g_A^\text{pre2002}$ differs from
$g_A^\text{post2002}$ by 5\sig. So, a case can be made that one should
not continue to include in averaging outlying values based on older
techniques when a significant disagreement arises. Indeed, the history
of $\tau_n$ and $g_A$ experimental shifts, indicates they come in
pairs as new technological methods emerge. In this case, the 2002
confirmation \cite{Abele:2002wc,Abele:1997yn} of a relatively large
$g_A$ with small errors may be viewed as the harbinger of a shorter
lifetime which several years later began to be directly observed in
trapped lifetime experiments. 

One might ask whether theory or some other weak interaction phenomenon
can be used to determine $g_A$ (and $\tau_n$ indirectly)?  On the
theory side there is the promise of lattice QCD
\cite{Capitani:2017qpc,Shintani:2016zvp}. The lattice approach is, in
principle, an ideally suited first principles method for computing a
relatively pure strong interaction effect such as $g_A$. However,
early lattice attempts to compute $g_A$ generally obtained smaller
than expected values with large systematic uncertainties. Recently,
the situation has been improving. Indeed, a recent study
\cite{Chang:2017oll} found the preliminary result
$g_A^\text{lattice}=1.285(17)$, in good agreement with our ``favored''
value at about the $\pm 1$\% level. How much further the lattice
precision can improve remains to be seen. Fortunately, the current
uncertainty is statistics dominated; so, long dedicated lattice
running can potentially reduce the error. Perhaps our suggestion of a
favored value with small uncertainty may help motivate a heroic
effort.

An alternative independent experimental $g_A$ determination using muon
capture in Muonic Hydrogen was recently examined
\cite{Hill:2017wgb}. Using theory and experimental input for other
parameters, the measured capture rate gave $g_A=1.276(11)$
i.e.~somewhat better than 1\% agreement with our favored value. It was
suggested in that study that a future factor of 3 improvement in the
measured capture rate combined with a better lattice determination of
the axial charge radius could provide a $g_A$ determination at the
level of $\pm 0.2$-0.3\%. That would be a nice check on our favored
values, but it would appear difficult to improve that
approach much further.

To illustrate an application of the favored values, we end our
discussion by deriving a general constraint on possible exotic (beyond
the standard model) neutron decays and applying it to an interesting
scenario recently put forward by Fornal and Grinstein (FG)
\cite{Fornal:2018sbb} in an effort to solve the neutron lifetime
puzzle. Those authors suggest that the BR=Branching Ratio for
radiative inclusive $n\to pe\bar\nu(\gamma)$ could be 0.99 rather than
1 due to a speculated 1\% exotic neutron branching ratio into dark
particle decay modes (e.g.~$n\to \text{dark }n+\text{scalar}$) without protons and electrons.  In that case beam
experiments that only detect decays with final state protons or
electrons would actually measure a partial lifetime
$\tau_n^\text{full}/\BR$ with $\BR<1$ while trapped neutron
experiments that count the number of neutrons as a function of time
measure the full inclusive lifetime $\tau_n^\text{full}$.  Although
throughout this work, we tacitly conclude that the beam lifetime is an
outlier whose value will shift in future more precise follow-up
experiments and eventually agree with our favored trapped
$\tau_n^\text{trap}=879.4(6)\second$, addressing the
Fornal-Grinstein solution is an instructive exercise that we will use
to conclude this paper.

We begin by generalizing our analysis to the case where the BR for
$n\to pe\bar\nu(\gamma)$ can be $< 1$ due to exotic decays such as
$ n\to $ dark particles.  In that case, eqs.~(\ref{eq:2}, \ref{eq:1}
and \ref{eq:5}) are modified by replacing $\tau_n$ with
$\tau_n^\text{full}/\BR$ where
$\tau_n^\text{full} = 1/(\text{total neutron decay rate})$.  That
replacement leads via eq.~\eqref{eq:5} to (assuming $\Vud$ extracted
from superallowed beta decays and CKM unitarity agreement are
negligibly affected by the exotic new physics)
\begin{equation}
  \label{eq:16}
\BR = \tau_n^\text{full}(1+3g_A^2)/5172.0(1.1)\second .
\end{equation}
Accepting $\tau_n^\text{trap}=879.4(6)\second$ as the full lifetime in eq.~\eqref{eq:16} and
expanding BR in $g_A$ about $g_A=1.2755$, the directly measured axial
coupling post2002 central value, leads to
\begin{equation}
  \label{eq:17}
        \BR = 0.9999(7)+1.30(g_A-1.2755)+\dots
\end{equation}
That formula demonstrates the closeness of $\BR$ to 1 for
$g_A^\text{favored}=1.2755(11)$. It suggests a degree of tension
between the recent determinations of $g_A^\text{post2002}$ and the
Fornal-Grinstein solution to the neutron lifetime puzzle.  In fact,
phrased as a one sided 95\% CL bound it requires
\begin{align}
  \label{eq:18}
                              1-\BR &= \text{Total Exotic Neutron Decay
                              Branching Ratio}\nonumber \\ & < 0.27\%
                            \text{ for } g_A=1.2755(11).
\end{align}
That bound implies that satisfying more than 2.4$\second$ of the
8.6$\second$ lifetime puzzle difference has less than a 5\% chance of
being realized. One can overcome such a likelihood restriction by
assuming a smaller $g_A$ in eq.~\eqref{eq:17}.  For example,
$g_A =1.268$ leads to $\BR=0.99$ which corresponds to about a
9$\second$ lifetime difference.  Any axial coupling roughly in the
range $1.268<g_A <1.272$ could account for a good part of the puzzle.
Unfortunately, there would be a price to pay for a smaller $g_A$ in that
range.  Those values are in disagreement with the most recent
$g_A^\text{post2002} = 1.2755(11)$ by 3 or more $\sigma$. Thus, the lifetime
puzzle would be replaced by a $g_A$ inconsistency.

The Fornal-Grinstein scenario will be tested by new measurements of
$\tau_n$ both beam and trap to see if the current puzzle survives and
needs a solution.  If so, the next step will be more precise
determinations of $g_A$ via neutron decay asymmetries or perhaps lattice
gauge theories. Will $g_A$ revert back to a smaller value?  Updates of
$g_A$ in the past have almost always led to larger values, but the past
is not always a good predictor for the future.

A scenario similar to that of Fornal and Grinstein was envisioned by
K.~Green and D.~Thompson \cite{Green:1990zz} for the rare decay
$n\to \text{hydrogen} + \bar\nu$.  They used the different effects of
that decay on beam and trap lifetimes to obtain a bound of $<3\%$ for
that branching ratio (to be compared with the $4\cdot 10^{-6}$
prediction
\cite{Bahcall:1961zz,Kabir:1967usj,McAndrew:2014iia,daudel:jpa-00234057}). Our
general analysis employing $\tau_n^\text{trap}$ and
$g_A^\text{post2002}$ in eq.~\eqref{eq:16} can be used to reduce that
bound by an order of magnitude to $<0.27\%$.

Our $0.27\%$ bound in eq.~\eqref{eq:18} also applies to neutron
oscillations into mirror or dark neutrons
\cite{Berezhiani:2005hv,Serebrov:2007gw,Serebrov:2018mva}, exotic
phenomena proposed to explain the neutron lifetime puzzle.

Future expected order of magnitude improvements in
$\tau_n^\text{trap}$ and asymmetry measurements, should improve the
sensitivity of our bound in eq.~\eqref{eq:18} to roughly
$3\cdot 10^{-4}$ for the exotic phenomena described above.

Acknowledgement: The work of A.~C.~was supported by the Natural
Sciences and Engineering Research Council of Canada. The work of
W.~J.~M.~was supported by the U.S. Department of Energy under grant
DE-SC0012704. The work of A.S. was supported in part by the National
Science Foundation under Grant PHY-1620039.
%

\end{document}